\def \yskip{\penalty-50\vskip3pt plus 3pt minus 2pt}
\def \reference{\par \yskip \noindent \hangindent .4in \hangafter 1}
\def \abc#1#2#3#4 {\reference#1, {\sl#2}, {\bf#3}, #4}
\def \blank {\lower 5pt\hbox to 0.75in{\hrulefill}}

\def \s{~\rm{s}}
\def \km{~\rm{km}}

\def \yrs{~\rm{yrs}}
\def \yr{~\rm{yr}}

\def \lesssim{\mathrel{<\kern-1.0em\lower0.9ex\hbox{$\sim$}}}
\def \gtrsim{\mathrel{>\kern-1.0em\lower0.9ex\hbox{$\sim$}}}

\documentstyle[11pt,aasms,tighten,flushrt]{article}

\begin{document}
\small

\setcounter{page}{1}



\title{Criticism of recent calculations of common envelope ejection}

\author{Noam Soker and Amos Harpaz}

\affil{Department of Physics, Oranim, Tivon 36006, Israel \\
soker@physics.technion.ac.il; phr89ah@techunix.technion.ac.il}


$$
$$

\centerline {\bf ABSTRACT}

We re-examine a recent claim by Han et al.\ (2002) that the ionization
energy in the envelope has to be included in the ejection criterion
of common envelopes.
In particular, we argue that (1) they include a mass loss rate prior
to the onset of the common envelope that is too low;
(2) They do not include the energy radiated by the accreting white dwarf
companion, as well as that emitted by the core of the giant star;
and (3) As argued by one of us before, the opacity in the envelope
is too low for the efficient usage of the ionization energy.

{\it Subject headings:} binaries: close
$-$ stars: evolution
$-$ stars: RGB and AGB
$-$ stars: mass loss

\section{INTRODUCTION}

In the common envelope (CE) phase a compact companion enters
the envelope of a more extended star, e.g., an asymptotic giant
branch (AGB) or red giant branch (RGB) star, and because of
tidal interaction and friction, the orbit shrinks (see reviews by
Iben \& Livio 1993 and  Taam \& Sandquist 2000).
A commonly used parameter is the ratio of the
binding energy of the ejected envelope, $\Delta E_{\rm bind}$, 
to the orbital energy that is released during the CE phase,
$\Delta E_{\rm orb}$:
$\alpha_{\rm CE} \equiv \Delta E_{\rm bind}/\Delta E_{\rm orb}$
(e.g., Livio \& Soker 1988; Sandquist et al. 1998;
Note that different definitions for the
binding energy exist, e.g., O'Brien, Bond, \& Sion 2001).
In some systems the usage of the above  expression in a simple
manner yields $\alpha_{\rm CE} >1$.
This led some researchers to argue that the energy stored in the
envelope, and in particular the ionization energy, i.e., the energy
released when the envelope material recombines, is the extra energy needed
to expel the CE (e.g., Han, Podsiadlowski, \& Eggleton 1994;
Dewi \& Tauris 2000; Maxted et al.\ 2002,
and earlier references in these papers).
 This proposed mechanism was criticized by us in previous papers.
Harpaz (1998) criticized the paper by Han et al.\ (1994), arguing that
after recombination the opacity drops sharply, hence the
released energy flows outward instead of pushing mass outward.
Soker (2002) criticized the paper by Maxted et al.\ (2002) for not
considering the mass lost from the envelope prior to the onset of the CE.
Eggleton (2002) notes that a close companion may substantially enhance
mass loss rate prior to the onset of a Roch lobe overflow (RLOF),
with the possibility of preventing a CE phase altogether.

In a very recent paper, Han et al.\ (2002; hereafter HA02), using
population synthesis to study the formation of subdwarf B stars
in binary systems, argue that ". . the ionization energy in the
envelope has to be included in the ejection criterion . .".
We do not agree with this assessment,  {{{ although we do agree
with their main results, that subdwarf B stars can be formed
through the channels they consider. }}}
Moreover, we are surprised to find that they do not confront the
criticism mentioned above, although they were acquainted with
its existence
{{{{ (in a later paper they briefly refer to our criticism;
Han et al.\ 2003). }}}}
This is a fundamental question in the CE process,
and it is relevant to many different kinds of close binary systems.
For example, Lobel et al. (2003) propose that the high mass loss rate
during the outburst of the yellow hypergiant $\rho$ Cassiopeiae
was driven by the release of hydrogen recombination energy.
For that, we critically examine some of the assumptions
and calculations in these papers.
In particular, we find the sections dealing with the CE
channel for the formation of subdwarf B stars in HA02,
to contain, what we consider, some flaws,
which we elaborate on in the next section, and which led
them to the above statement regarding the ionization energy.

\section{THE CRITICISM}

\subsection {The enhanced mass loss rate due to fast rotation.}
In their simulations, HA02 start the CE phase when
the primary RGB star overflows its Roche lobe.
Using the approximate equation for Roche lobe size (Eggleton 1983),
we find that for most cases they explore RLOF that takes place when
$R_g \sim 0.5 a$, where $R_g$ is the giant radius,
and $a$ is the the orbital separation.
Because of tidal interaction, the secondary will bring the giant
envelope to corotate with the orbital motion when the RGB stellar
radius becomes $R_g \gtrsim 0.15 a$ (see scaled equations in Soker 1996).
We {{{ first bring a qualitative treatment, and }}}
examine the last $\sim 30 \%$ of the radius evolution,
namely from $R_g \simeq 0.35 a$ to $R_g \simeq 0.5 a$.
During this period the orbital velocity of the synchronizely-rotating
RGB star increases from $\omega \simeq 0.2 \omega_{\rm Kep}$
to $\omega \simeq 0.35 \omega_{\rm Kep}$, where $\omega_{\rm Kep}$
is the (critical) Keplerian  angular velocity on the giant's equator.
Such a high rotation velocity of a deeply convective envelope is
likely to result in strong magnetic activity, which most likely
enhances the mass loss rate (e.g., Soker \& Clayton 1999).
If a fraction $\eta = 0.1$ of the radiation momentum is invested in
blowing a slow wind at a speed $v=10 \km \s^{-1}$, then by
momentum balance we find the mass loss rate to be
$\dot M = 2 \times 10^{-7} (L_g/10^3 L_\odot) M_\odot \yr^{-1}$,
where $L_g$ is the stellar luminosity.
This is $\gtrsim 10$ times larger than the rate assumed by HA02.
For several million years of evolution during this stage, the RGB star
may lose an extra mass of few tenth to one solar mass. 
We note that in AGB stars $\eta$ can become much larger,
with $\eta \sim 1$, and with a higher mass loss rate for the
same stellar luminosity.
Our view that RGB stars rotating at $\omega \sim 0.1 \omega_{\rm Kep}$
can lose substantial fraction of their envelope, is discussed along
with supporting arguments for low mass RGB stars in globular clusters,
in Soker \& Harpaz (2000).
Eggleton (2002) argues for enhanced mass loss rate from supergiants
stars close to filling their RLOF.

{{{ We now try a more quantitative approach.
A commonly used expression for the enhanced mass loss rate in binary
systems is
\begin{equation}
\dot M_g = A_1 \frac {R_g L_g}{M_g}
\left[1 + B_L \left( \frac {R_g}{R_L} \right)^\gamma \right]
=A_1 \frac {R_g L_g}{M_g} (1+Q),
\end{equation}
where $M_g$ is the giant's mass, and $R_L$ is the radius of a sphere
that has the same volume as the Roche lobe.
The second equality defines $Q$, which depends most strongly
on $R_g$.
Here and in the rest of this section $A_i$, $i=1,2,3 . .$, $B$,
$B_L$ and $\gamma$ are constants.
Different values for $B_L$ and $\gamma$ are quoted in the literature:
Tout \& Eggleton (1988) use $B_L=10^4$, $\gamma=6$, and if
$R_g/R_L>0.5$ then they set 0.5 for this ratio.
Han et al.\ (1995) prefer $B_L \simeq 500$ and $\gamma=6$,  
while Han (1998) argue fore $B_L \simeq 1000$ and $\gamma=6$.  
Frankowski \& Tylenda (2001) argue for a more complicated
expression, which basically has a very low value of $B_L$ and $\gamma=3$;
however, their numerical calculations yield a much faster increase
in the mass loss rate as the giant is close to fill its Roche lobe.
Since we mainly refer to HA02, we take the lower
range used by these authors, i.e., $\gamma=6$ with $B_L=500-1000$.
Therefore, the maximum value of $Q$ is
$Q_{\rm max}=B_L/2^6 \simeq 8-16$.

To follow the evolution of an RGB star, we use the analytical
approximate relations from Iben \& Tutukov (1984)
for RGB stars which are descendant of population I
main sequence stars in the mass range $0.8 < M/M_\odot < 2.2$.
These read
\begin{equation}
\frac {R_g}{R_\odot} =10^{3.5}
\left( \frac {M_c}{M_\odot} \right)^4 \qquad {\rm and} \qquad
\dot M_c  =10^{-5.36}
\left( \frac {M_c}{M_\odot} \right)^{6.6},  
\end{equation}
where $M_c$ is the core mass.
The luminosity is proportional to $\dot M_c$, and from the last equation
can be written as
\begin{equation}
L_g =  A_2 R_g ^ {6.6/4}.
\end{equation}
The time interval is evaluated from equation (2) 
\begin{equation}
dt= \frac {d M_c}{\dot M_c} =
A_4 \frac {R_g^{-3/4} d R_g }{R_g^{6.6/4}} =
A_4 R_g^{-9.6/4} d R_g. 
\end{equation}
Substituting for $L_g$ and $d t$ from equations
(3) and (4), respectively, in equation (1) yields
\begin{equation}
M_g d M_g = A_5 R_g^{1/4} (1 + B R_g^\gamma) d R_g.  
\end{equation}
For constant values of $A_5$, $B$, and $\gamma$, this can be integrated.
Assuming that the final radius is much larger than the radius at
which synchronization is achieved, and taking $M_{g0}$ to
be the initial giant's mass, we find
\begin{equation}
M_{g0}^2-M_g^2 = \frac {8}{5} A_5
R_g^{5/4} \left(1 + \frac {5}{5+4 \gamma} B R_g^\gamma \right).
\end{equation}
Taking the mass that was lost to be $\Delta M_g = M_{g0}-M_g$,
and using the definition of $Q$ and $Q_{\rm max}$ given in
equation (1), gives for the mass lost prior to the onset of the CE
\begin{equation}
\Delta M_g = \frac {4}{5} A_5 \frac{R_g^{5/4}}{M_{g0}}
\left(1 + \frac {5}{5+4 \gamma} Q_{\rm max} \right)
+\frac {(\Delta M_g)^2}{2 M_{g0}}.
\end{equation}

We find now the approximate ratio between the mass lost by an RGB
star in a binary system before entering the CE phase,
and the mass lost by a single RGB star.
For that, we assume that in both cases the maximum radius is the same.
This is reasonable concerning that the subdwarf B stars required
that the giant went through a helium flash. 
We also ignore the last term in equation (7); including it
will increase this ratio, hence will increase the
magnitude of the effect studied here.
For a single RGB star $Q=0$, and we find
\begin{equation}
Z \equiv \frac {\Delta M_g({\rm binary})}{\Delta M_g({\rm single})}
\gtrsim 1 + \frac {5}{5+4 \gamma} Q_{\rm max}.
\end{equation}
For $\gamma=6$ and the values range $Q_{\rm max} \simeq 8-16$
quoted after equation (1) above, we find $Z \gtrsim 2.3-3.7$.

With this ratio in hand, we return to the calculations of HA02.
We consider their case with $\alpha_{CE}=0.5$,
i.e., half the released gravitational energy of the
spiraling-in companion goes to expel the envelope;
no ionization energy is assumed in this case
(case b in their Sec. 3.4.1).
We examine the distribution of final orbital separations
as given in their figures (4a), with no mass loss, and (4c), with
a coefficient of 0.5 in the Reimers' wind mass loss rate.
We find that the tail of this distribution at large orbital
separations is larger by a factor of $\sim 1.3-1.4$ in the case
with wind mass loss;
namely, the required gravitational energy in the case with wind mass
loss is $\simeq 1/1.35 \simeq 0.75$ times that in the case with
no wind mass loss.
This crudely implies that in the case with wind mass loss the
companion enters the envelope when the envelope mass is
$M_{\rm env} \simeq 0.75 M_{\rm env0}$, where $M_{\rm env0}$
is the envelope mass when no wind mass loss is considered.
Using our estimation for the increase in the envelope mass lost
because of binary interaction, $Z \sim 2.3-3.7$, we estimate that  
when the enhanced mass loss rate because of binary interaction
is considered, the total envelope mass lost will be
$\Delta M_{\rm env} = 0.5-0.9 M_{\rm env0}$.
This implies that the final orbital separations in the
large orbital separation range will increase by a factor of
$\sim 2-10$ relative to the case with no wind mass loss.
This is an increase by a factor of $\sim 3-30$ in the
orbital periods.
Taking this into account in the graph of the orbital periods
distribution for $\alpha_{\rm CE}=0.5$ in figure (4a) of HA02,
will lower the peak of this graph and extend it to larger
orbital periods.
This brings it much closer to agreement with the observed distribution.
It seems that even for lower values of $\alpha_{\rm CE}$
the inclusion of the enhanced mass loss rate can bring
the results closer to the observed distribution, without the inclusion
of ionization energy.
}}}

{{{{ We conclude that Han et al.\ (2002) use of mass loss rate
$\leq 0.5$ times Reimers' mass-loss rate (Reimers 1975),
is inappropriate, while also ignoring enhanced mass
loss rate due to fast rotation, tidal interaction, and other effects
of binary interaction.
The enhanced mass loss rate expected of rapidly rotating RGB and
AGB stars is an important factor not to take into account. }}}}

It should be noted that the scenario outlined here
is somewhat different from that in Soker (2002), although the
basic arguments concerning $\alpha_{\rm CE}$ are the same.
Here the RGB star loses a substantial fraction of its envelope
as it expands by a relatively large factor from the moment of
synchronization to the RLOF.
Soker (2002) mainly considers AGB stars which have much higher mass
loss rates, hence they do not expand much after synchronization
takes place and before they lose their entire envelope;
they still can expand a little, mainly
because of thermal pulses (helium flashes).
The CE occurs as angular momentum is lost in the intensive AGB wind,
and for a narrow range of parameters the companion
enters the AGB envelope after a substantial fraction of the envelope
have been lost; again, reducing the required value of $\alpha_{\rm CE}$.

\subsection {Inclusion of the energy gained from the accreting WD.}
The accretion rate of a WD orbiting inside a CE is limited by the
Eddington luminosity.   This accretion can be a significant energy source
{{{ (Iben \& Livio 1993; }}} Armitage \& Livio 2000).
Han et al.\ (2002) use either a WD of mass $M_{WD}=0.3 M_\odot$
or $M_{WD}=0.6 M_\odot$.
 The total energy radiated at the Eddington luminosity during a time
$\tau_{eq}$ will equal the amount of energy liberated when
a mass $M_e$ recombines when
\begin{equation}
\tau_{eq} \simeq 10
\left( \frac {M_e}{1 M_\odot} \right)
\left( \frac {M_{WD}}{0.6 M_\odot} \right)^{-1}
\yr.
\end{equation}
Since the spiraling-in process is expected to take longer than
100 years (see fig. 10 of Han et al.\ 2002), the total energy liberated
by the accreting WD is much larger than that emitted by the recombining
envelope.
Even the total energy radiated by the RGB core,
with $L \simeq 1,000 L_\odot$, becomes larger than the ionization energy
after $\sim 200 (M_e/1 M_\odot) \yrs$, a period which is 
shorter than the spiraling-in process of a WD of mass
$M_{WD} = 0.3 M_\odot$ (fig. 10 of Han et al.).
As argued by Soker (2002), an energy source is not 
a restriction in ejecting a CE;
it is the momentum balance that requires careful consideration.

The situation may be different during outbursts in massive
yellow hypergiant stars, e.g., $\rho$ Cassiopeiae, where a large
fraction of the envelope is retained during the outburst.
Therefore, the ionization energy released by a large mass in the
envelope can contribute to the ejection of a small fraction of the
envelope (Lobel et al.\ 2003).

\subsection {Realistic opacity consideration.}
Consider a radiation of total energy $E$ which is
accelerating a wind to velocity $v$.
The total momentum supplied by the wind is $\beta E/c$, where
$c$ is the light speed and $\beta$ is the effective mean number of
times a photon is scattered in the wind before it escape.
In AGB stars, which lose mass at a very high rate, typically
$\beta \lesssim 3$ (Knapp 1986).
 Let us consider the case in which the recombination energy,
converted to radiation, effectively expels a fraction $f$ of the
recombining mass to a wind's speed $v$.
This requires the effective mean number of times a photon is scattered
to be
\begin{equation}
\beta \sim 200
\left( \frac {f}{0.1} \right)
\left( \frac {v}{10 \km \s^{-1}} \right).
\end{equation}
 This is a huge number of times, which requires a very high opacity in
the wind.
But as argued previously by one of us (Harpaz 1998), the recombination
sharply reduces the opacity, making the radiation emitted inefficient
in ejecting the CE, in particular in the inner regions
(Sandquist, Taam, \& Burkert 2000).
More than that, as evidenced from planetary nebulae with binary nuclei,
most of the ejected CE mass is in the equatorial plane
(Bond \& Livio 1990).
Hence radiation will more easily escape along the polar direction.
The requirement on the envelope-opacity value in order to facilitated
the ionization energy becomes unrealistically high.

{{{
\section {SUMMARY}

In a recent paper HA02 study the formation of subdwarf B stars
in binary systems.
We basically agree with their main results, but criticized here
their claim that
". . the ionization energy in the envelope has to be included
in the ejection criterion . .".
In Section 2.1 we estimate the final orbital separation after a common
envelope phase, using the distribution of binary systems found
by HA02 in their population synthesis, and with enhanced mass loss
rate because of binary interaction, as the same authors used in previous
papers. 
We find that if a reasonable enhancement in the mass
loss rate is included, the final orbital distribution becomes
much closer to the observed distribution, without the inclusion
of ionization energy.
In sections 2.2 and 2.3 we reexamine two other effects which
strengthen our claim; these are the energy supply by the
accreting companion which may help in expelling the envelope
(Iben \& Livio 1993), and the problem in using the ionization energy
as opacity drops with recombination (Harpaz 1998). }}}

\acknowledgements
{{{ We thank Adam Frankowski for useful discussions. }}}
This research was supported by the US-Israel Binational Science
Foundation and the Israel Science Foundation.


\begin{references}

\reference{} Armitage, P. J., \& Livio, M. 2000, ApJ, 532, 540

\reference{} Bond, H. E., \& Livio, M. 1990, ApJ, 355, 568

\reference{} Dewi, J. D. M., \& Tauris, T. M. 2000, A\&A, 360, 1043

\reference{} Eggleton, P. P. 1983, ApJ, 268, 368

\reference{} Eggleton, P. P. 2000, ApJ, 575, 1037

\reference{} {{{ Frankowski, A., \& Tylenda R. 2001, A\&A, 367, 513 }}}

\reference{} {{{ Han, Z. 1998, MNRAS, 296, 1019 }}}

\reference{} {{{ Han, Z., Eggleton, P. P., Podsiadlowski, Ph., \&
           Tout, C. A. 1995, MNRAS, 277, 1443 }}}
            
\reference{} Han, Z., Podsiadlowski, Ph., \& Eggleton, P. P. 1994,
  MNRAS, 270, 121

\reference{} {{{{ Han, Z., Podsiadlowski, Ph., Maxted, P. F. L.,
            \& Marsh, T. R., 2003, MNRAS, in perss (astro-ph/0301380) }}}}
            
\reference{} Han, Z., Podsiadlowski, Ph., Maxted, P. F. L., Marsh, T. R.,
 \& Ivanova, N. 2002, MNRAS, 336, 449 (HA02)

\reference{} Harpaz, A. 1998, ApJ, 498, 293

\reference{} Iben, I. Jr., \& Livio, M. 1993, PASP, 105, 1373

\reference{} {{{ Iben, I. Jr., \& Tutukov, A. V. 1984, ApJSupp, 54, 335 }}}

\reference{} Knapp, G. R. 1986, ApJ, 311, 731

\reference{} Livio, M., \& Soker, N. 1988, ApJ, 329, 764

\reference{} Lobel, A., Dupree, A. K., Stefanik, R. P., Torres, G.,
    Israelian, G., Morrison, N., de Jager, C., Nieuwenhuijzen, H.,
    Ilyin, I., \& Musaev, F. 2003, ApJ, 583, 923 

\reference{} Maxted, P. F. L., Burleigh, M. R., Marsh, T. R.,
      \& Bannister, N. P. 2002, MNRAS, 334, 833
        
\reference{} O'Brien, M. S., Bond, H. E., \& Sion, E. M. 2001,
            ApJ, 563, 971

\reference{} Reimers, D., 1975, Mem. Roy. Soc. Liege 6e Ser, 8, 369

\reference{} Sandquist, E. L., Taam, R. E., \& Burkert, A.
      2000, ApJ, 533, 984

\reference{} Sandquist, E. L., Taam, R. E., Chen, X.,
    Bodenheimer, P., \& Burkert, A. 1998, ApJ, 500, 909

\reference{} Soker, N. 1996, ApJ, 460, L53

\reference{} Soker, N. 2002, MNRAS, 336, 1229

\reference{} Soker, N., \& Clayton, G. C. 1999, MNRAS, 307, 993

\reference{} Soker, N. \& Harpaz, A. 2000, MNRAS, 317, 861

\reference{} Taam, R. E., \& Sandquist, E. L. 2000, ARA\&A, 38, 113

\reference{} {{{ Tout, C. A., \& Eggleton, P. P. 1988, MNRAS, 231, 823 }}}

\end{references}
\end{document}